\documentstyle[11pt]{article}
\newcommand{\blankline}{\vskip .3cm}
\newcommand{\f}{\begin{equation}}
\newcommand{\ff}{\end{equation}}

\begin{document}
\centerline{\LARGE   A holographic formulation of quantum general
relativity }
\blankline
\rm
\centerline{Lee Smolin${}^*$}
\blankline
\centerline{\it  Center for Gravitational Physics and Geometry}
\centerline{\it Department of Physics}
\centerline {\it The Pennsylvania State University}
\centerline{\it University Park, PA, USA 16802}
\vfill
\centerline{Revised October 21, 1999}
\vfill
\centerline{ABSTRACT}
We  show that there is a sector of quantum general relativity, in the
Lorentzian signature case, 
which may be expressed in a completely holographic formulation in
terms of states and operators defined on a finite boundary.
The space of boundary states is built out of the conformal blocks of 
$SU(2)_L \oplus SU(2)_R$,  $WZW$ field theory 
on the $n$-punctured sphere, where $n$ is related to the
area of the boundary.  The Bekenstein bound is explicitly  satisfied.  
These results are based on a new lagrangian and hamiltonian formulation of
general relativity based on a constrained $Sp(4)$ topological
field theory.  The hamiltonian formalism is polynomial, and also left-right 
symmetric.
The quantization uses balanced $SU(2)_L \oplus SU(2)_R$ spin networks and 
so justifies the state sum model of Barrett and Crane. By extending the 
formalism to $Osp(4/N)$ a holographic formulation of extended
supergravity is obtained, as will be described in detail in a
subsequent paper.
\vfill
\blankline
${}^*$ smolin@phys.psu.edu
\eject

\section{Introduction}

There has been recently much interest in holographic formulations
of theories of quantum gravity\footnote{This is a revised version of
the original preprint hep-th/9808191.  Besides setting right a minor
error the main change is that a more general boundary condition is 
given for the Lorentzian signature theory, which has an infinite
dimensional solution space.  Thanks are due to Yi Ling
for discussions during the course of joint work on the supersymmetric
extension, to appear \cite{superholo}.}
Besides the original argument based on the 
Bekenstein bound
of 't Hooft\cite{thooft} and Susskind\cite{lenny-holo}, there is also a 
very interesting argument based on
results of topological quantum field theory advocated by 
Crane\cite{louis-preholo} and others\cite{pluralistic,lotc,relational} that 
suggests
that quantum cosmological theories should be based on states and observables
living on boundaries inside the universe.  These two arguments reinforce
each other in an interesting way: the Bekenstein bound\cite{bek} tells us that 
there should be a finite amount of information per unit area of the boundary 
while
topological quantum field theories provides a large class of quantum field
theories with finite dimensional state spaces associated to boundaries.  

For these reasons, several years ago a holographic formulation of quantum
general relativity was presented\cite{linking}. The theory was holographic in
that the physical state space had the explicit form,
\f
{\cal H}_{\cal B}= \sum_a {\cal H}_a
\ff
where $a$ are the eigenvalues of the area operator $\hat{A}$
(which is known by both 
construction\cite{spain,vol1,sn1} and general theorems\cite{gangof5}
to have a discrete spectra.)  The eigenspaces of definite area were
constructed explicitly in terms of the conformal blocks of $SU(2)_q$
$WZW$ conformal field theory on the punctured two sphere.  
More explicitly, the areas are expressed in terms of a sums of total
quantum spins $j_i$ associated with the punctures, so that in the large $k$
limit\cite{qdef}
\f
a(j_i) = \sum_i l_{Pl}^2 \sqrt{j_i (j_i+1)}
\ff

\f
{\cal H}_{a(j_i)} = {\cal V}_{j_i} 
\ff
where ${\cal V}_{j_i}$ is the space of conformal blocks (or intertwiners)
on the punctured two sphere.
  
It then follows from the formula for the dimension of these spaces that 
the Bekenstein bound\cite{bek} is satisfied, so that\cite{linking}
 \f
dim({\cal H}_{A}) \leq e^{c \over 4 G_B \hbar}
\label{bek}
\ff
where $c=\sqrt{3} /Ln(2)$ in quantum general relativity and $G_B$ is the
``bare" Newton's constant.  Thus, this result implies
that the macroscopic Newton's constant, which is not so far
predicted by the theory, should be $G= G_B/c$.

Finally a complete set of boundary observables based on the gravitational
fields at the boundary exists that is both sufficient to make complete
measurements of the physical state and expressed explicitly in terms
of operators in the conformal field theory\cite{linking}.

Another property of this formulation is that the bulk state which describes the
physics in the interior of the boundary is the Chern-Simons state of
Kodama\cite{kodama}, which is known to have a semiclassical interpretation
in terms of de Sitter or Anti DeSitter spacetime\cite{kodama,chopinlee}.

 These results show that, at least for quantum general relativity, completely
holographic formulations exist.  

Given the recent interest in holographic 
formulations of $\cal M$ 
theory
\cite{matrix,horava,juan-ads,witten-ads,other-ads}, it is 
then
very natural to try to extend these results to ${\cal N}=8$ supersymmetry, to 
provide
a candidate for a completely background independent formulation of $\cal M$
theory.  This goal was the impetus of the present work.  However, in order
to accomplish the supersymmetric extension, certain issues had to be addressed, 
which led to a new formulation of general relativity at  both the classical 
and quantum
level.
As these may be of independent interest, they are presented here.
A subsequent paper will
presents  an extension of the present results
to theories with extended supersymmetry, some of which may
be candidate for such a formulation of
$\cal M$ theory, in a $3+1$ dimensional compactification\cite{n=8}

The new formulation presented here is related to the 
Ashtekar formulation \cite{abhay,action,sam}, but 
differs
from it in that it is entirely left-right symmetric.  Both self-dual and 
antiself-dual fields
are kept in the theory, although they are in the end related to each other 
through 
constraints that play the role of the reality conditions.  At the same time, the
formulation is entirely polynomial\footnote{Another way of modifying
the Ashtekar formalism that uses two connections is given in
\cite{barbero}.}.  

This formulation has several features that are of interest for holographic 
formulations
of the theory. First, because the reality conditions are part of the algebra of 
constraints, 
the lorentzian form of the theory is more easily studied.  Second, the extension 
to 
the supersymmetric case is somewhat easier, as will be seen in the 
subsequent
paper.  
Third,  it allows a more transparent treatment of the splitting between
kinematical and dynamical constraints, in both the bulk and boundary theories.

This last point is the most important and is worth elaborating on.  The basic 
idea
of the formalism is that general relativity is expressed as a constrained
topological field theory, for the group $G=Sp(4)$.  This group, which double
covers the Anti-deSitter group contains $H=SU(2)_L \oplus SU(2)_R$ as a subgroup.
What is meant by a constrained topological field theory is that all derivative
terms, and hence the structure of the canonical theory, is the same as a 
topological field theory with no local degrees of freedom.  The local degrees of
freedom arise because of the imposition of 
local, non-derivative constraints, which reduce the
explicit gauge symmetry from $G$ to the subgroup $H$.  
The fields in the coset $G/H$ become the gravitational degrees of
freedom, further, the constraints in the coset become non-linear and
in fact become the generators of spacetime diffeomorphisms.

What is interesting is that to extend to the case of $\cal N$-supersymmetry, all
that is needed is to extend the structure just described so that 
$G=Osp(4/{\cal N})$ and the subgroup $H$ is some supersymmetric extension
of $SU(2)_L \oplus SU(2)_R$,
with at most half the supersymmetry generators of $G$.  This, and several
related ideas, are discussed in \cite{n=8}.  In this paper we
describe the classical and quantum physics of the non-supersymmetric theory.

\section{General relativity as a constrained TQFT}

In this section we introduce
new way of writing general relativity as a constrained topological quantum
field theory, which we call the ambidextrous formalism
\footnote{We may note that there is more than one way to represent general
relativity with a cosmological constant 
as a constrained topological quantum field theory.  The earliest
such approach to the authors knowledge is that of 
Plebanski \cite{pleb}, studied also in \cite{rl}.  Alternatively,
one can deform a topological field theory of the form of
$\int Tr F\wedge F$, as described in \cite{CDJ} 
(see also \cite{spain}).. What is new in the present presentation is
the representation of general relativity as a constrained topological
field theory for the DeSitter group $SO(3,2)$.  For reasons that will
be apparent soon, the present formulation is more suited both to
the Lorentzian regime and to the theory with vanishing cosmological
constant.}.
 For the non-supersymmetric case we study here, the theory is based
on a connection valued in the Lie algebra
$G=Sp(4)$, (which double covers $SO(3,2)$ the anti-deSitter group.)
Thus, this approach is similar to that of MacDowell-Mansouri,
in which general relativity is found as a consequence of breaking
the $SO(3,2)$ symmetry of a topological quantum field theory
down to $SO(3,1)$\cite{mm}. However it differs from that approach in
that the beginning point is a $\int B\wedge F$ theory.

The $Sp(4)$ connection is written $A_{\alpha \beta}$ where the four dimensional
indices $\alpha, \beta,=1,...,4$ will be often broken down into
a pair $\alpha = (A , A^\prime)= (0,1,0^\prime,1^\prime)$ of 
$SU(2)$ indices expressing
the fact that $SU(2)_L \oplus SU(2)_R \subset Sp(4)$. 
Thus, the connection is
\f
A_{\alpha \beta}= \{ A_{AB},{A}_{A^ \prime B^\prime} , 
A_{AA^\prime} \}.
\ff
The components of the connection $A_{AA^\prime}$ which parameterize
the coset
$Sp(4)/SU(2)_L \oplus SU(2)_R$ will be taken to represent the frame
fields  $e_{AA'}$ so we will take
\f
A_{AA^\prime}={1\over l}e_{AA^\prime}
\ff
where $l$ has dimensions of length.

We take for our starting point a modification of the
 $Sp(4)$ $B \wedge F$ theory.  This is given by 
\f
I^0=  \imath  \int_{\cal M} {1 \over g^2} \left (  
B_{\alpha}^{\ \beta} \wedge F_{\beta}^{\ \gamma}\gamma_{5 \gamma}^{\ \alpha}
\right )
-{e^2  \over 2} \left (
B_{\alpha}^{\ \beta} \wedge B_{\beta}^{\ \gamma}\gamma_{5 \gamma}^{\ \alpha} 
\right )
+{-\imath k \over 4\pi} \int_{\partial {\cal M}} ( Y_{CS}(A_{AB}) -
Y_{CS}(A_{A'B'}) )
\label{theaction}
\ff
where $B_{\alpha}^{\ \beta}$ is a two form valued in the adjoint 
representation
of $Sp(4)$, $Y_{CS}$ is the $SU(2)$ Chern-Simons action, $g$ and
$e$ are 
dimensionless coupling constants and $k$ is as usual an integer.
The variational principle given by (\ref{theaction}) is well defined
only in the presence of certain boundary conditions, which are the 
subject of the next section.  

$\gamma_{5}$ is given by
\f
\gamma_{5 \alpha}^{\beta} = \left ( 
\begin{array}{cc}
	\delta_{A}^{\ B} & 0  \\
	0 & -\delta_{A'}^{\ B'}
\end{array}
   \right ) 
\ff

The inclusion of the $\gamma_{5}$ is necessary if we want the 
action to be parity invariant\footnote{Note that this is not
required to reproduce classical general relativity, as this can
be done with parity asymmetric actions\cite{action}.  However,
we insist on it here as we want to develop a form of the quantum
theory which is explicitly parity invariant.  It is interesting to
note that (\ref{theaction} remains an action for general
relativity if the $\gamma_{5}$ is replaced by $\delta_{\alpha}^{\beta}$
In this case the action is chiral but the $SP(4)$ gauge symmetry
is broken only by the constraints.}.
However, its presence breaks the 
$SP(4)$ invariance down to $SU(2)_{L} \oplus SU(2)_{R}$.
To see this we expand to find
\begin{eqnarray}
I^0 & = &  -\imath  \int_{\cal M} {1 \over g^2} \left (
B_{AB} \wedge F^{AB} 
- B_{A'B'} \wedge F^{A'B'}
\right ) \nonumber \\
&& - {e^2  \over 2} 
  \left (
B_{AB} \wedge B^{AB} 
- B_{A'B'} \wedge B^{A'B'}
\right )  \nonumber \\
&&+{\imath k \over 4\pi} \int_{\partial {\cal M}} \left ( Y_{CS}(A_{AB}) -
Y_{CS}(A_{A'B'}) \right )
\end{eqnarray}
Thus we see that we have $-\imath$ times 
the difference between the actions for the
$\int B\wedge F $ theories for $SU(2)_L$ and $SU(2)_R$.  
The mixed components $B^{AA'}$ have disappeared from the
theory.  The
reason for preferring this choice will be clear shortly.

We now impose
two constraints that set the $SU(2)_L \oplus SU(2)_R$ components
of $B^{\alpha \beta}$ to be equal to the self-dual and antiself-dual
two forms constructed from $e^{AA^\prime}$.  With constraints
that do this the action has the form,
\begin{eqnarray}
I^1 &=& -\imath \int_{\cal M}  {1 \over g^2 } \left (  B^{AB}\wedge F_{AB}
- B^{A^\prime B^\prime }\wedge F_{A^\prime B^\prime}   \right ) 
- {e^2  \over 2} 
\left (
B_{AB} \wedge B^{AB} 
- B_{A'B'} \wedge B^{A'B'}
\right ) 
\nonumber \\
&&+
\lambda_{AB} \wedge \left (  
{1 \over l^2} e^{AA^\prime}\wedge e^B_{\ A^\prime} - B^{AB}  \right )
+ \lambda_{A^\prime B^\prime } \wedge \left (  
{1 \over l^2} e^{A^\prime A}\wedge e^{B^\prime}_{\ A} - B^{A^\prime B^\prime} 
\right ) \nonumber \\
&& +{\imath k \over 4\pi} \int_{\partial {\cal M}} \left ( Y_{CS}(A_{AB}) -
Y_{CS}(A_{A'B'}) \right )
\label{I1}
\end{eqnarray}

It is not hard to show that the equations of motion of this
action reproduce those of general relativity with a cosmological
constant.  To see this we note the forms of the $Sp(4)$ curvatures,
\f
F^{AB}=f^{AB}+ {1 \over l^2} e^{A A^\prime}\wedge e^B_{\ A^\prime}
\label{FAB}
\ff
\f
F^{AA^\prime}={\cal D}e^{AA^\prime}
\ff
where $f^{AB}$ is the $SU(2)_L$ curvature of the connection
$A_{AB}$ and ${\cal D}$  is the $SU(2)_L \oplus SU(2)_R$
covariant derivative.  $F^{AA'}$ is the torsion.
(The definition of $F^{A^\prime B^\prime}$
is the same as (\ref{FAB}) with primed indices.)
The $\lambda_{AB}$ and $\lambda_{A^\prime B^\prime}$ field equations 
set
\f
B^{AB}= {1 \over l^2} e^{AA^\prime}\wedge e^B_{\ A^\prime}   \equiv {1 \over 
l^2} \Sigma^{AB}
\label{phi}
\ff
\f
B^{A^\prime B^\prime} = {1 \over l^2} e^{A^\prime A}\wedge e^{B^\prime}_{\ A} 
\equiv {1 \over l^2}
\Sigma^{A^\prime B^\prime}  
\label{phibar}
\ff
Putting the solutions to these field equations
back into the action, we find,
\begin{eqnarray}
I^1 &=& -\imath \int_{\cal M} 
{1\over G}  \left ( e^{AA^\prime}\wedge e^B_{\ A^\prime} \wedge f_{AB}
-e^{A^\prime A}\wedge e^{B^\prime}_{\ A} \wedge f_{A^\prime B^\prime}
\right ) \nonumber \\
&&+ \Lambda \  e^{A A^\prime}\wedge e^B_{\ A^\prime} \wedge
e_{AB^\prime}\wedge e_B^{\ B^\prime} 
\nonumber \\
& & +{\imath k \over 4\pi} 
 \int_{\partial {\cal M}} \left ( Y_{CS}(A_{AB}) -
Y_{CS}(A_{A'B'}) \right )
\end{eqnarray}
where 
\f
G= g^2 l^2
\ff
and
\f
\Lambda = { 2 \over l^4}\left  ( {1 \over g^2} - {e^2 \over 2}   \right )
\label{cosmo}
\ff
This is an action for general relativity in first order form.
The reason for the funny signs and factors of $\imath$ is that
\f
e^{a} \wedge e^{b} \wedge e^{c}\wedge e^{d}\epsilon_{abcd}
= (-\imath) \left ( \Sigma^{AB}\Sigma_{AB} -\Sigma^{A'B'}\Sigma_{A'B'}
\right )
\ff
with a similar identity holding for the curvature term.

To show the complete
correspondence with general relativity we may consider 
the $A_{AB}$ and $A_{A'B'}$ field equations  which,
(ignoring the
boundary terms) are,
\f
{\delta I^1 \over \delta A_{AB}} : D \wedge B^{AB}=0
\label{AAB}
\ff
\f
{\delta I^1 \over \delta A_{A'B'}} : \bar{D} \wedge B^{A'B'}=0
\label{AA'B'}
\ff
Together with (\ref{phi}) and (\ref{phibar}) these
give,
\f
{\cal D} \wedge \Sigma^{AB}= {\cal D} \wedge \Sigma^{A'B'}=0
\ff
which implies that the $SU(2)_L \oplus SU(2)_R$ connections $A_{AB}$
$A_{A'B'}$ are the metric compatible torsion free connections associated
with the frame fields $e^{AA'}$.  This in turn implies that the torsion,
\f
F_{AA'} = \nabla \wedge e_{AA'} =0.
\ff
The action is then,
\begin{eqnarray}
I^1 &=& -\imath \int_{\cal M} 
{1\over G}  \left ( e^{AA^\prime}\wedge e^B_{\ A^\prime} \wedge f_{AB}[A(e)]
-e^{A^\prime A}\wedge e^{B^\prime}_{\ A} \wedge f_{A^\prime B^\prime}[A(e)]
\right ) \nonumber \\
&&+ \Lambda  \ e^{A A^\prime}\wedge e^B_{\ A^\prime} \wedge
e_{AB^\prime}\wedge e_B^{\ B^\prime} 
\nonumber \\
&&
+{\imath k \over 4\pi} 
 \int_{\partial {\cal M}} \left ( Y_{CS}(A_{AB}) -
Y_{CS}(A_{A'B'}) \right )
\end{eqnarray}
which is Einstein's action with a cosmological constant.

Equivalently, we
may plug the solutions to the constraints into the remaining
field equations to find the field equations for general relativity.
The complete set of field equations are (\ref{phi}), (\ref{phibar}),
(\ref{AAB}) and (\ref{AA'B'})
together with,
\f
{\delta I^1 \over B^{AB} }= {1\over g^2} F_{AB} - e^2 B_{AB} - \lambda_{AB}=0,
\label{BAB}
\ff
its double with primed indices everywhere and 
\f
{\delta I^1 \over e^{AA^\prime}} : 
e^B_{\ A^\prime} \wedge \left (
{1 \over l^2} B_{AB} + \lambda_{AB}   \right )
+ e^A_{\ B^\prime} \wedge \left (
{1 \over l^2} B_{A'B'} + \lambda_{A'B'}  
\right )  =0
\label{e}
\ff
Plugging (\ref{BAB}) and its primed double into
equation (\ref{e}) we then find the Einstein equation.
\f
{1 \over G}\left (  f^A_B \wedge e^{BA'}  + f^{A'}_{B'} \wedge e^{AB'} 
\right ) - \Lambda e^{AB'} \wedge e^B_{B'} \wedge e^{A'}_B =0
\ff
Thus, we have shown how general relativity with
a cosmological constant may be derived
as a constrained $Sp(4)$ $B\wedge F$ theory.

We may note that because the action contributes two terms to
the cosmological constant, there is the possibility of canceling
the cosmological constant, while preserving the structure which
derives from an $SP(4)$ connection.  From (\ref{cosmo})
we see that $\Lambda =0$ for
\f
e^2 = {2 \over g^2} 
\label{zerocosmo}
\ff
This is interesting as it implies that the cosmological constant
vanishes at a kind of self-dual point.

This derivation has also shown that there is some redundancy
in the field equations that follow from (\ref{I1}).   
As is known from \cite{action} the left and right handed field
equations decouple so the right handed part of the connection
can be gotten instead from the left handed part of the connection
by imposing reality conditions, $A_{A'B'}=\bar{A}_{AB}$.

Thus, {\it as far as the bulk equations of motion are concerned} we
can further constrain the $Sp(4)$ symmetry of the $B\wedge F$ theory
down to only $SU(2)_L$ by setting $B^{A'B'}=B^{AA'}=0$ to find
the bulk action
\begin{eqnarray}
I^2 &=& {1 \over g^2 } \int_{\cal M} \left (  B^{AB}\wedge F_{AB}
+ \lambda_{AB}\left (  
e^{AA^\prime}\wedge e^B_{\ A^\prime} - B_{AB}  \right )
\right )  \label{Isd}
\end{eqnarray}

In the Euclidean case this self-dual action suffices, while in the
Lorentzian case it must be supplemented by the reality condition,
$A_{A'B'}=\bar{A}_{AB}$.  It will be important to keep this in mind
when we turn to the study of the boundary theory.

\section{The role of the boundary terms in the field equations}

When we take the boundary conditions into account we must impose
some boundary conditions to insure that the action (\ref{I1}) is functionally
differentiable.  The equations we need to worry about are the
$A_{AB}$ and $A_{A'B'}$ field equations.  When we
make a variation the boundary contributes a term of the form
\f
\delta I^1_{boundary}    = 
\imath \int_{ \partial {\cal M}}   \left [  \delta A_{AB} \wedge  ( 
{k \over 4\pi } f^{AB} -{1 \over g^2 l^2} \Sigma^{AB} ) 
- \delta A_{A'B'} \wedge  ( 
{k \over 4\pi } f^{A'B'} -{1 \over g^2 l^2} \Sigma^{A'B'} ) \right ].
\label{bt}
\ff
In order for the action to be functionally differentiable we must then impose
a boundary condition that makes (\ref{bt}) vanish.  

There are several different boundary conditions that might be
imposed.  We will be interested here in a set of boundary conditions
that extend the ``self-dual boundary conditions''
studied, in the case of the euclidean theory, in \cite{linking}.
These were motivated by the fact that they allow 
DeSitter of AntideSitter spacetime to be solutions. 
In the Lorentzian theory we can impose similar conditions, but
the details are different, as we now describe\footnote{I am grateful
to Abhay Ashtekar, Yi Ling and Roger Penrose for discussions on
these conditions.}

In the Euclidean case we imposed in \cite{linking} the condition that 
the pullbacks of the fields into the boundary satisfied the pull
back of the self-dual equations, expressed on two forms as 
\f
\vec{f}^{AB} = {4 \pi \over l^{2} g^{2} k}  \vec{\Sigma}^{AB} .
\label{sdsd}
\ff  
where $\vec{f}$ indicates the pull back of the two forms into the boundary. 
These are of course satisfied
by DeSitter or Antidesitter spacetime, as the full two forms satisfy
these conditions.  However, in the Euclidean case there are an 
infinite number of other spacetimes whose two forms pulled back to the
boundary satisfy (\ref{sdsd}).  This 
is because the left and right handed parts of the curvature are 
independent in the Euclidean case. As a result, the Euclidean theory
with (\ref{sdsd}) imposed on the boundary has a solution space given
by one degree of freedom for each point on the boundary.  This may
be verified explicitly by linearized analysis\cite{linking}.

In the Lorentzian case the left and right handed parts of the Weyl
curvature are not independent, they are complex conjugates of each 
other.  Hence we cannot impose (\ref{sdsd}) and have $f^{A'B'}$
vary independently on the boundary, as in the Euclidean case.  In
fact, in that case it can be verified that the result of the reality
conditions is to limit the freedom in the solutions to (\ref{sdsd})
to oscillations of the boundary in deSitter or Antidesitter spacetime.
To have an infinite dimensional space of classical solutions in the
Lorentzian case we must relax the boundary conditions.   

To see how this may be done, we note first 
that   (\ref{bt}) can be made to vanish in two ways.  We can either
fix the connection on the boundary and require that $\delta \vec{A}$ vanishes 
or we can require that the self-dual conditions (\ref{sdsd}) be 
satisfied.
However it is also possible to study mixed conditions in which
we choose the first solution for some components of $\vec{A}$
and the second for the other components.   

One thing we would like to retain in the Lorentzian case is the
relationship between the dimension of the state space of the boundary
theory and the area of the boundary, discovered in \cite{linking}
as this provides a realization of the Bekenstein bound\cite{bek}.
However, as we will see when we study the canonical quantization,
this only requires that the self-dual boundary conditions (\ref{sdsd})
be imposed on the pull back of the two forms into the intersections of
the boundary and the spatial slices.  For the other components
we can relax the boundary conditions.  One natural way to do this
is the following\footnote{For more details concerning the application
of these boundary conditions, see \cite{superholo}.}.

First, we fix a time slicing of the boundary $\partial {\cal M}$.  
We choose a time
coordinate, $t$,  such that these are $t=constant$ slices.  $t$
is then fixed up to a one parameter time reparametrization group
$t \rightarrow t^{\prime}= f(t)$. 
We will then impose the self-dual condition (\ref{sdsd})
on the $t=constant$ spatial slices of $\partial {\cal M}$.

We want to weaken the boundary conditions by imposing (\ref{sdsd})
on only some of the mixed space-time components of the boundary.
We can do this locally by fixing coordinates $\sigma^{1},\sigma^{2}$
on the $t=constant$ slices of $\partial {\cal M}$.  We then 
fix the self dual boundary conditions for the following components,
\f
\vec{f}^{AB}_{\sigma^{1}\sigma^{2}} =   {4\pi \over g^2 kl^{2}}  
\vec{\Sigma}^{AB}_{\sigma^{1}\sigma^{2}}
\label{boundary1}
\ff
\f
\vec{f}^{A'B'}_{\sigma^{1}\sigma^{2}} =   {4\pi \over g^2 kl^{2}}  
\vec{\Sigma}^{A'B'}_{\sigma^{1}\sigma^{2}}
\label{boundary1'}
\ff
and
\f
\vec{f}^{AB}_{t\sigma^{1}} =   {4\pi \over g^2 kl^{2}}  
\vec{\Sigma}^{AB}_{t\sigma^{1}}
\label{boundary2}
\ff
\f
\vec{f}^{A'B'}_{t\sigma^{1}} =   {4\pi \over g^2 kl^{2}}  
\vec{\Sigma}^{A'B'}_{t\sigma^{1}}
\label{boundary2'}
\ff
However, we make the remaining terms in (\ref{bt}) vanish
by putting 
\f
\delta A_{\sigma^{1}}^{AB} = \delta A_{\sigma^{1}}^{A'B'}=0
\label{boundary3}
\ff
Clearly these conditions are compatible with the reality conditions,
and they result in a functionally differentiable action.  At the same 
time the fact that the self-dual conditions (\ref{sdsd}) are not
imposed on all pull-backs of the two forms on the boundary means
that the solution space is larger.  While $A_{1}$ is fixed, there is now no
condition on $\dot{A}_{\sigma^{2}}$. Consequently
that  component of the connection is allowed to evolve, so long
as (\ref{boundary1},\ref{boundary1'}) are satisfied.

The rationale for these conditions comes from the quantum theory, more
particularly from the form of the
boundary Hilbert space, which is constructed as in \cite{linking}
from the spaces of intertwiners of the quantum deformed gauge
group.  In the quantum theory we define the boundary conditions by a
condition that the spin networks intersect the boundary at a discrete
set of points, called punctures, whose labels do not evolve.
This does constrain certain components
of the connection (up to local gauge transformations). The reason is 
that the traces of the holonomies
around a loop $\gamma$ that surrounds a single puncture are fixed
by the quantization of the conditions 
(\ref{boundary1},\ref{boundary1'}).  Thus, fixing certain components
of the connection on the boundary is a consequence of fixing the
boundary conditions in the quantum theory in such a way that the
labels on the punctures that determine the Hilbert space of the
boundary theory are fixed and do not evolve.  But by doing so we
weaken the boundary conditions on other components of the connection. 
This gives the boundary state more freedom to evolve within those
fixed Hilbert spaces.
We will see how this works when we come to the quantum theory in 
section 6.

To complete the specification of the boundary conditions we 
will then anticipate the role of the punctures in the quantum theory
and fix a discrete  set of preferred points on the
spatial boundary.  Each such puncture is surrounded by a local
region and in each of these we may  introduce
local coordinates $(r,\theta)$ which are angular coordinates
with the puncture at the origin.  These can then be joined yielding
a single coordinate patch on the whole punctured sphere, which
reduces to an angular coordinate system in the neighborhood of
each puncture. Bringing back $t$ we then have a coordinate
system $(r,\theta, t)$ on the whole of ${\partial M}$ minus
the world lines of the punctures.  We then apply the above
conditions with $\theta=\sigma^{1}$ and $r=\sigma^{2}$.
The boundary conditions (\ref{boundary3})  then imply that the holonomies of the 
$SU(2)_{L} \oplus SU(2)_{R}$ connections around loops in the 
spatial boundary that surround
single punctures are fixed.

Finally note that compatibility of
(\ref{boundary1},\ref{boundary1'}) with the field equations  requires that
\f
G\Lambda =  
{8\pi \over l^{2}g^2 k}  
\label{r1}
\ff
which gives us a relation
\f
k= {4 \pi \over 1- {g^2 e^2\over 2}}
\label{r2}
\ff
This is an interesting relation, as $k$ must be an integer.
We see that at the self dual point where $G^2 \Lambda =0$,  
$k \rightarrow \infty$.

\section{The canonical formalism}

To understand the relationship between the 
$Sp(4)$ gauge invariance and diffeomorphism invariance, as well
as to prepare to discuss the quantization we study in this section
the canonical formulation of the theory we have just introduced.
We do this by making a $3+1$ decomposition of
the action (\ref{I1}) in the usual way\cite{action}, with the spacetime
manifold decomposed as ${\cal M}=\Sigma \times R$, with 
$\Sigma$ a three manifold.  Here we ignore boundary terms; their effects are 
included in 
the following section.

Before beginning, we must fix a point of view concerning the relationship
between the complex quantities such as the self-dual connections 
and the real metric of spacetime.  We will take here the approach in which
all fields are assumed to be complex and then set up the canonical 
formalism
for this case.  We will then consider the reality conditions to be a 
restriction
on the space of solutions which is 
imposed after the canonical formalism is set up.  This is natural for 
considerations of the quantization, because it parallels the situation of
the quantum theory in which the operator algebra is defined over the 
complexes, while the reality conditions are imposed by the choice of an
inner product.  At the level of the abstract algebra, before the inner
product is imposed, it makes no sense to restrict to the real sector, as 
this
is done by restricting certain operators to be hermitian, but this is not 
defined in the absence of the inner product.

We now proceed to the $3+1$ decomposition.  We write the action in terms
of space and time coordinates separately, with spacetime index 
$\mu=(0,a)$, with $a=1,2,3$, we have,
\begin{eqnarray}
I&= &-\imath \int dt \int_{\Sigma} \epsilon^{abc} \{
{1 \over g^2} B^{AB}_{ab} \dot{A}_{cAB} - 
{1 \over g^2} B^{A'B'}_{ab} \dot{A}_{cA'B'}     \nonumber \\
&&+{1 \over g^2} A^{AB}_0 [ {\cal D}_a B_{bcAB } ]
-{1 \over g^2} A^{A'B'}_0 [ {\cal D}_a B_{bcA'B' }  ]  \nonumber \\
&&+ e_{AA'0}  [{2 \over g^2 l}
B_{ab}^{AB} e_{cB}^{\ \ A'} 
+{2\over l^2} \lambda^{AB}_{ab} e_{cB}^{\ \ A'} -
{ 2 \over g^2 l^2} B^{A'B'}_{ab} e_{cB'}^A -
{2 \over l^2} \lambda^{A'B'}_{ab} e_{cB'}^{\ \ A} ]  \nonumber \\
&&+B_{0a}^{AB} [ {1\over g^2} f_{bcAB} 
+ {1 \over g^2 l^2} e_{bA}^{\ \ A'} e_{cBA'} -
e^2  B_{bcAB} - \lambda_{bcAB}    ] \nonumber \\
&&- B_{0a}^{A'B'} [ {1\over g^2} f_{bcA'B'} 
+ {1 \over g^2 l^2} e_{bA'}^{\ \ A} e_{cB'A} -
e^2  B_{bcA'B'} - \lambda_{bcA'B'}    ] \nonumber \\
&&+ \lambda_{0a}^{AB} [{1 \over l^2} e_{bA}^{\ \ A'} e_{cBA'}
- B_{bcAB} ]
  \nonumber \\
&&+ \lambda_{0a}^{A'B'} [{1 \over l^2} e_{bA'}^{\ \ A'} e_{cB'A}
-B_{bcA'B'}  ]  \} 
\end{eqnarray}
The canonical momenta for the forms $B$, $\lambda$, and
$e_{0AA'}$ all vanish, as do the canonical momenta of
the time components $A_{0AB}, A_{0A'B'}$.
This gives the  primary constraints.
The nonvanishing canonical momenta are for $A_{aAB}$ and $ A_{aA'B'}$
are, respectively,
\begin{eqnarray}
\pi^{a}_{AB} &=&  {-\imath \over g^2}  \epsilon^{abc} B_{bcAB}  \nonumber \\
\pi^{a}_{A'B'} &=&  {\imath \over g^2}  \epsilon^{abc} B_{bcA'B'}  \nonumber \\ 
\end{eqnarray}
The $\pi^a$'s are, as usual, vector densities.

We now come to the secondary constraints.  First there are the
$SU(2)_L \oplus SU(2)_R$ gauge constraints, which are,
\begin{eqnarray}
G^{AB} &=&{\cal D}_a \pi^{aAB}  =0 \\
G^{A'B'} &=& {\cal D}_a \pi^{aA'B'} =0
\end{eqnarray}
These preserve the vanishing of the canonical momenta of
$A_{0AB}$ and $A_{0A'B'}$.  The vanishing of the canonical
momenta for $e_{0AA'}$ is more complicated, and gives the
four secondary constraints,
\f
G^{AA'} =  
{1 \over l^2} [ \pi^{cAB}- \imath \lambda^{*cAB} ]  e_{cB}^{\ \ A'}   
+ {1 \over l^2} [ \pi^{cA'B'} -  \imath
\lambda^{*cA'B'} ]  e_{cA}^{\ \ B'}  =0
\ff
One might expect that as the $Sp(4)/(SU(2)_L \oplus SU(2)_R)$
gauge symmetry seems to be explicitly broken by the
constraints in the action, these would become second class
constraints.  Instead, as we shall see, these four equations
become the hamiltonian and diffeomorphism constraints of the
theory.

From the vanishing of the canonical momenta for the mixed space-time
components of the two forms
$B_{0a}^{AB}$ and $B_{0a}^{A'B'}$we get two more sets of constraints, 
\begin{eqnarray}
I^{aAB} &=& \epsilon^{abc} \left [ {1\over g^2}
f_{bc}^{AB} +{1 \over g^2 l^2} e_{b}^{AA'} e_{cA'}^{\ \ B} - 
 e^2 B_{bc}^{AB} -  \lambda_{bc}^{AB}
\right ]  \\
I^{aA'B'} &=& \epsilon^{abc} \left [{1\over g^2}
f_{bc}^{A'B'} +{1 \over l^2} e_{b}^{A'A} e_{cA}^{\ \ B'} - 
 e^2 B_{bc}^{A'B'} +  \lambda_{bc}^{A'B'}
\right ] 
\end{eqnarray}
The first pair, $I^{aAB}$ and $I^{aA'B'}$ may be solved to express
the $\lambda_{ab}^{AB}$ and $\lambda_{ab}^{A'B'}$ in terms of
the other fields.  These constraints are then eliminated with the
primary constraints which are the vanishing of the $\lambda$'s
momenta.

The preservation of the vanishing of the canonical momenta for
the mixed components $\lambda_{0a}^{AB}$ and
$\lambda_{0a}^{A'B'}$ results in six more constraints
that show that  the $\pi^{aAB}$ and $\pi^{aA'B'}$ are fixed
to be the duals of the self-dual two forms constructed from the
frame fields:
\begin{eqnarray}
J^{aAB}&=&\pi^{aAB}+ {\imath \over g^2 l^2} 
\epsilon^{abc}e_b^{AA'}e_{cA'}^{\ \ B} =0
\\
J^{aA'B'} &=& \pi^{aA'B'}- {\imath \over g^2 l^2} 
\epsilon^{abc} e_b^{A'A} e_{cA}^{\ \ B'} =0
\label{piaA'B'}
\end{eqnarray}
The  can be solved to eliminate the
$e_a^{AA'}$ in terms of the $\pi^{aAB}$ and the quantities
$N_{AA'}$.  These are four quantities defined by
\f
N_{AA'}= t^\mu e_{\mu AA'}
\ff 
where $t^\mu$ is the timelike
unit normal \cite{action}.  They are subject to the one constraint
$N^{AA'}N_{AA'}=2$, which follows from $t^\mu t_\mu = -1$.
They therefor represent three independent quantities, which
together with the nine $\pi^{aAB}$ allow us to express the
twelve $e_a^{AA'}$ as,
\f
e_a^{AA'}= e_{aA}^{\ B}  N_B^{\ A'} = { 1 \over \sqrt{q}}\epsilon_{abc} 
\pi^{bBC} 
\pi^{cA}_{\ \ C} N_B^{\ A'}
\label{edef}
\ff
We also have the complex conjugate of these relations
\f
e_a^{A'A}= e_{aA'}^{\ B'}  N_{B'}^{\ A} = { 1 \over 
\sqrt{\bar{q}}}\epsilon_{abc} 
\pi^{bB'C'} \pi^{cA'}_{\ \ C} N_{B'}^{\ A}
\label{edef'}
\ff
where $q$ and $\bar{q}$ are made from the determinants of
$\pi^{AAB}$ and $\pi^{aA'B'}$ in the usual way.

We may note that even for the complex case, the
$\pi^{aAB}$ and $\pi^{aA'B'}$ are not independent quantities.
This is because
the pullback of the self-dual and antiself-dual three forms 
of a metric define the same metric.  As a result there is
an additional second class constraint, which is
\f
R^{ab}=\pi^{aAB} \pi^b_{AB} - \pi^{aA'B'} \pi^b_{A'B'}=0
\label{Rab}
\ff
We will come back to the role this plays after we have
isolated the hamiltonian constraint of the theory.

For completeness we mention also two
more sets of constraints, which express the
lagrange multiplier fields in terms of other quantities.  They
they play no role in what follows as the lagrange multipliers 
are in any case eliminated but we give them for completeness.

The preservation of the vanishing of the momenta for the
$\lambda_{ab}$'s result in constraints
\begin{eqnarray}
J_a^{AB}&=& B_{0a}^{AB}- {2 \over l^2} e_{0A'}^{\ A} e_{a}^{\ A' B} =0\\ 
J_a^{A'B'}&=& B_{0a}^{A'B'}- {2 \over l^2} e_{0A}^{\ A'} e_{aA'A}  =0
\end{eqnarray}

Similarly, the preservation of the vanishing of the momenta for the
$B_{ab}$'s result in constraints
\begin{eqnarray}
I_{aAB}&=& - {1 \over g^2} 
{\cal D}_a A_{0AB} + {1 \over g^2} \dot{A}_{aAB} +
{1 \over g^2 l^2} e_{0AA'}e_{aB}^{\ \ A'}-2 e^2  B_{0a}^{AB}
+ \lambda_{0a}^{AB} =0\\
I_{aA'B'}&=& - {1 \over g^2}  {\cal D}_a A_{0A'B'} + {1 \over g^2} 
\dot{A}_{aA'B'} +
{1 \over g^2 l^2} e_{0A'A}e_{aB'}^{\ \ A}-2e^2  B_{0a}^{A'B'}
+ \lambda_{0a}^{A'B'} =0
\end{eqnarray}
Using all the $I$ and $J$ constraints, we find that $G^{AB}$ and
$G^{A'B'}$ are unchanged, and indeed are first class constraints
that generate $SU(2)_L \oplus SU(2)_R$ internal gauge
transformations.  However, the components of the Gauss law
in the coset $Sp(4) / SU(2)_L \oplus SU(2)_R$  now become,
\f
G^{AA'}= {1\over l^2} e_{cB}^{\ \ A'} \left [
{\Lambda g^{2 }\over l^{2}} \pi^{cAB} 
-{2\imath \over g^{2}} \epsilon^{abc} f_{bc}^{AB}
\right ] + {1\over l^2} e_{cB'}^{\ \ A} \left [
{\Lambda g^{2 }¥\over l^{2}} \pi^{cA'B'} 
+ {2\imath \over g^{2} }\epsilon^{abc} f_{bc}^{A'B'}
\right ] 
\ff
where the cosmological constant is defined by (\ref{cosmo}).

We now use (\ref{edef}) and (\ref{edef'}) to write these in terms of six 
new constraints,
\f
G^{AA'}= {9 G \over N^{2} E^{3}} \left ({ N^{A'}_D \over \sqrt{q}} C^{AD} 
+ {N^{A}_{D'} \over \sqrt{q} }C^{A'D'} 
\right )
\label{combo}
\ff
where,
\begin{eqnarray}
C^{AD} &=& -\imath  \pi^{aC}_{\ B} \pi^{bD}_{\ C} f_{ab}^{AB}  +
{ \Lambda g^{4 } \over l^{2}}
\epsilon_{abc} \pi^{aC}_{\ B} \pi^{bD}_{\ C} \pi^{cAB} \\
C^{A'D'} &=& \imath \pi^{aC'}_{\ B'} \pi^{bD'}_{\ C'} f_{ab}^{A'B'}  + 
{ \Lambda g^{4 } \over l^{2}}
\epsilon_{abc} \pi^{aC'}_{\ B'} \pi^{bD'}_{\ C'} \pi^{cA'B'} 
\label{hams}
\end{eqnarray}

These constraints must vanish independently because they transform
separately under $SU(2)_L$ and $SU(2)_R$ transformations.  Thus,
from the closure of the constraint algebra we have
\f
\{ G^{AB} , G^{CC'} \} \approx C^{AB} \approx 0 
\ff
\f
\{ G^{A'B'} , G^{CC'} \} \approx C^{A'B'} \approx 0
\ff

Now we must recall that the two conjugate pairs $(A_{aAB}, \pi^{aAB})$
and $(A_{aA'B'}, \pi^{aA'B'})$ are  mutually commuting, so that,
\f
\{C^{AB} , C^{A'B'} \} =0
\ff
Furthermore, we know from work of Jacobson in \cite{tedopen} that 
the four $C^{AB}$ contain the standard Hamiltonian and diffeomorphism
constraints of the Ashtekar formalism, and thus make a first class 
algebra.
The same is then true for the $C^{A'B'}$.  It follows that the algebra
of the four $G^{AA'}$ is first class and contains the hamiltonian and
diffeomorphism constraints of the theory.  To see this in more detail,
consider the four vector $V^{AA'}=V^\mu e_{\mu}^{AA'}$ 
as a parameter of the constraints
\f
G(V)= \int V^{AA'}G_{AA'} = \int W^{AB}C_{AB} + W^{A'B'}C_{A'B'}
\ff
where for simplicity we have set 
\f
W^{AB}= {9G \over N^{2} E^{3}} {N^{A'}_B \over \sqrt{q}} V^{AA'}
\ff
and similarly for $W^{A'B'}$.  We have thus expressed $G(V)$ in
terms of lagrange multipliers and two copies of the
Ashtekar constraints.  Thus, their algebra is first class.  It
also follows that the algebra of $G^{AA'}$ with both
$G^{AB}$ and $G^{A'B'}$ is first class.  

FInally, we must deal with the
is the remaining constraint $R^{ab}$ given by
(\ref{Rab}).  Its Poisson bracket
with $G(V)$ gives a remaining set of constraints, which are
\begin{eqnarray}
S^{ab}(x) &\equiv & \{ R^{ab} (x) , G(V) \}  \nonumber \\
&=& {\cal D}_c ( W_A^D \pi^{c \ E}_B \pi^{(b}_{D)E} ) \pi^{aAB}
+ {\cal D}_c ( W_{A'}^{D'} \pi^{c \ E'}_{B'} \pi^{(b}_{D')E'} ) \pi^{aA'B'}
=0 .
\label{Sab}
\end{eqnarray}

This is actually a well known condition, it is the reality condition
for the Ashtekar formalism, which guarantees that
$\dot{q}^{ab}$ is real.  Here it is recovered as a constraint, even in
the complexified case.  It implies a relationship between the
real parts of $A_{aAB}$ and $A_{aA'B'}$.

In fact we can now give a simple interpretation of the resulting 
formalism.
With all fields complex, what we have are two copies of the Ashtekar
formalism, one with positive chirality and one with negative chirality.
However, the left and right sectors are related by the constraints
$R^{ab}$ and $S^{ab}$ that require that all metric quantities constructed
from the left and right handed sectors agree.
Given that the constraints come in the combination $G^{AA'}$ given by
(\ref{combo}) we have only four spacetime constraints, so the two copies
of the Ashtekar formalism evolve together with common lapses and shifts.
Thus, as in the Ashtekar formalism, once one sets the constraints
$R^{ab}$ and $S^{ab}$ to be zero, they are preserved in time, so that
the metric quantities continue to agree, whether computed from the
left or right sector.
Finally, even in the presence of the constraints,  
the internal gauge constraints are independent, so that the
local gauge symmetry is $SU(2)_L \oplus SU(2)_R$.  

Finally, so far we have not made a restriction to real metrics.
To do so is simple, we restrict to the subspace of the solution
space for which $q^{ab}$ and its time derivative are real.  Given the 
relations just found the equivalence to the Ashtekar formalism guarantees 
that real initial data will evolve to a real spacetime.

\section{The boundary theory in the canonical formalism}

We now include in the canonical analysis the effects of the boundary 
term in the action, 
proportional to the Chern-Simons invariant of the pull back of 
the connection on 
the boundary.
This analysis was first done in the chiral formulation in
\cite{linking}, here we extend it to the ambidextrous formulation.

With the boundary terms included, the primary constraints that define the 
non-vanishing momenta are,
\f
S^{a AB } (x) \equiv \pi^{aAB} (x) + {\imath \over g^2} 
B^{*a AB } 
(x)
 -{ \imath k \over 4 \pi} \int d^2 S^{ab} (\sigma )A_b^{AB}
\delta^3(x, S(\sigma))  =0
\label{primary}
\ff
\f
S^{a A'B' } (x) \equiv \pi^{aAB} (x) - {\imath \over g^2} 
B^{*a A'B' } 
(x)
 +{ \imath k \over 4 \pi} \int d^2 S^{ab} (\sigma )A_b^{A'B'}
\delta^3(x, S(\sigma))  =0
\label{primary'}
\ff
What is important for the construction of the boundary theory is the 
interaction
of the boundary term in the definition of the momenta (\ref{primary})
and the generalized Gauss's law constraints that come from the 
$A_0^{\alpha \beta}$ 
field equations. Recall that the 
Gauss's law for $SU(2)_{L} \oplus SU(2)_{R}$ t has the form,   
\f
G^{AB} \equiv{\imath \over g^{2}} {\cal D}_a B^{*aAB} 
\label{GG}
\ff
\f
G^{A'B'} \equiv {\imath \over g^{2}} {\cal D}_a B^{*aA'B'} 
\label{GG'}
\ff

If we use the definition of the momenta from (\ref{primary},\ref{primary'})
in the Gauss's law we find, after integrating by parts,
that
\begin{eqnarray}
G(\Lambda ) & \equiv & \int_\Sigma \Lambda_{AB}
G^{AB}   
= \int_{\Sigma} \Lambda_{AB} 
{\imath \over g^{2}} {\cal D}_a B^{*aAB}  \nonumber \\
&=& -\int_\Sigma 
{\cal D}_a (\Lambda_{AB}) \pi^{aAB } 
+\int_{\partial \Sigma} d^2S^{ab} \Lambda_{AB}
\left (
 {\imath k \over 4 \pi} f_{ab}^{AB}-\pi^{*AB}_{ab}
\right )
\label{BG}
\end{eqnarray}
with an identical expression for $G^{A'B'}$.
Thus, in addition to the bulk constraints we found in the
previous section, there are two boundary constraint given
by 
\begin{eqnarray}
G^B ( \lambda ) &=&
\int_{\partial \Sigma} d^2S^{ab} \lambda_{AB}
\left (
\pi^{*AB}_{ab} - {\imath k \over 4 \pi} f_{ab}^{AB}
\right )  
\label{bGAB} \\
\bar{G}^B ( \bar{\lambda} ) &=&
\int_{\partial \Sigma} d^2S^{ab} \bar{\lambda}_{A'B'}
\left (
\pi^{*A'B'}_{ab} - {\imath k \over 4 \pi} f_{ab}^{A'B'}
\right )  
\label{bGA'B'}  
\end{eqnarray}
These implement eqs (\ref{boundary1},\ref{boundary1'}),
which were the spatial parts of the boundary conditions
we  imposed to make the action functionally differentiable.  

The next thing to notice is  that the boundary 
term in the primary constraints (\ref{primary},\ref{primary'})  have
the effect of  modifying the Poisson brackets
for fields pulled back into the boundary.  We can see this
by computing their algebra.  Defining
$S(f) \equiv \int_\Sigma f_{a A B} S^{a AB}$,
we find,
\f
\{ S(f) , S(g) \} = {\imath k \over 2\pi} \int_{\partial \Sigma}
d^2S^{ab}f_{a AB} g_{b}^{AB}  .
\label{sc}
\ff
with similar relations holding for the equations with the
primed indices.

We can now characterize the kinematics of the boundary theory
classically.  The phase space of the boundary theory,
which we will call $\Gamma^{\partial \Sigma}$
can be characterized by fields pulled back to the
spatial boundary, which are
written $\vec{A}^{AB}_a, \vec{A}^{A'B'}_a$,
$\vec{\pi}^{*AB}_{ab}$ and $\vec{\pi}^{*A'B'}_{ab}$.
(Note that for these pullback fields the abstract
indices $a,b,c...$ are two dimensional.)
The latter commute with all other boundary fields
and hence
label sectors of the boundary phase space (They fail
to commute with connection variables normal to the boundary,
which are not part of the phase space of the boundary theory.)
  
By the constraints 
$\pi^{*AB}_{ab}$ and $\pi^{*A'B'}_{ab}$ are determined
(up to the $SU(2)_L \oplus SU(2)_R$
gauge invariance) in terms of the two metric on the boundary.
 
The actual degrees of freedom of the boundary phase
space are given by the 
$SU(2)_L \oplus SU(2)_R$ connection $A_a$, pulled back
into the boundary.  To find their Poisson brackets one
must construct the Dirac brackets by inverting the
second class constraints (\ref{sc}).  This is done
in detail in \cite{linking}, the result is
\f
\{ \vec{A}^{AB}_a (\sigma ) , \vec{A}_{bCD} 
(\sigma' ) \} = {2\pi \over k} \epsilon_{ab} \delta^2 (\sigma \sigma')
\delta^{(AB)}_{CD}
\ff
These are in fact the Poisson brackets of two dimensional
Chern-Simons theory.  

However, the curvatures of the boundary connection
are determined by the boundary terms in the
Gauss's law (\ref{BG}).  These require,
\begin{eqnarray}
\vec{f}^{AB} &=& {4 \pi \over k l^2} 
\vec{e}^{AC'}\wedge \vec{e}^B_{\ C'}  \nonumber \\
\vec{f}^{A'B'} &=& {4 \pi \over k l^2} 
\vec{e}^{A'C}\wedge \vec{e}^{B'}_C
\end{eqnarray}
There are relations between
$\vec{f}^{AB}$ and $\vec{f}^{A'B'}$.
These follow from the constraints which
express the fact that the pull backs of the self-dual
and anti-self-dual two forms into the spatial boundaries
$\Sigma$ define the same two geometry.  
These require that the invariants constructed from
$\vec{A}^{AB}$ and $\vec{A}^{A'B'}$ must be equal.

Thus, the phase space of the boundary theory is that of
$SU(2)_L \oplus SU(2)_R$ Chern-Simons theory, with an
external field constraining the curvatures.

The Hamiltonian of the theory may be constructed, following
the standard procedure, by extending the Hamiltonian constraint
by a boundary term so that the expression is functionally
differentiable even when the lapse function is non-vanishing
on the boundary.  To extract the Hamiltonian we may choose
$W^{AB} = \tau \epsilon^{AB}$ and $W^{A'B'} = \tau \epsilon^{A'B'}$.
The Hamiltonian then must have the form,
\f
H(\tau) = \int_{\Sigma} \tau \left [ \epsilon^{AB} C_{AB} + 
\epsilon^{A'B'} C_{A'B'} \right ]
+ \int_{\partial \Sigma} \tau h
\ff
where we require that the time coordinate $\tau$ match the 
slicing of the boundary given by the preferred $t=constant$ surfaces 
that go into the definition of the boundary conditions.  This means 
that continued to  the boundary $\tau$ must be a function
of $t$ which is constant on the $t=constant$ surfaces.   

The condition that $H$ be functionally differentiable
requires that $h$ be a functional defined on the boundary, of 
the form,
\f
\int_{\partial \Sigma} \tau h  = 4\imath \int_{\partial \Sigma} 
d^{2} S_{a} \tau \left [ \pi_{A}^{a B}  \pi_{B}^{b C}A_{bC}^{\ A}
-\pi_{A'}^{a B'}  \pi_{B'}^{b C'}A_{bC'}^{\ A'}
 \right ]
 \label{boundham}
\ff
When the constraints are satisfied this last expression,
(\ref{boundham}) is the Hamiltonian of the theory.  We see
that it is a functional on the boundary, which is both as required
by diffeomorphism invariance and consistent with the holographic
hypothesis.

\section{Quantization}

We may now sketch the quantization of the ambidextrous theory. 
We only emphasize those aspects which differ from the treatment
given for the Euclidean signature theory in \cite{linking}, to which
the reader may refer for more details.
We
begin with the bulk theory and then add the boundary
degrees of freedom.

We work first in the connection representation.  Initially the
configuration space is defined to be the space of complexified
$SU(2)_L \oplus SU(2)_R$ connections, mod 
internal gauge transformations:
\f
{\cal C}^{gauge} = {  (A^{AB} , A^{A'B'} )
\over G^{AB} \times G^{A'B'}   } .
\ff
Functionals on ${\cal C}^{gauge}$ will live in a
Hilbert space, subject to a suitable norm such
as that given in \cite{reiner,gangof5} called ${\cal H}^{gauge}$.

We must now discuss a subtle but important issue, having to do with
the use of the spin network states to describe the Lorentzian 
signature theory.  
For the case that the gauge group is real $SU(2)_L \oplus SU(2)_R$ 
the resulting space
of states has a basis given by the spin networks, as discussed
in \cite{sn1,gangof5}.  In these states the edges of the spin networks
are labeled by pairs of integers $j_{L},j_{R}$ corresponding
to the finite dimensional representations of $SU(2)_L \oplus SU(2)_R$.
In the present case, where the spacetime is Lorentzian the 
connections actually live in the complexification of $SU(2)_L \oplus SU(2)_R$.
This means that there is additional freedom in the choice of 
states, arising from the fact that the gauge group is non-compact.
One might choose, for example, to label the spin networks with 
continuous as well as discrete labels, corresponding to the full
set of representations of the gauge group.  

The strategy guiding the present approach is to set up a
quantization of the complexification of general relativity at the
kinematical level, and then impose the reality conditions as
operator equations, by realizing (\ref{Rab}) and (\ref{Sab})
on a suitable space of states.  Thus, in principle we do have the
freedom to work within a kinematical state space which is considerably
enlarged from that defined in \cite{sn1,gangof5} by extending the
labels on the spin networks to all representations of the 
complexifications of $SU(2)_L \oplus SU(2)_R$.  
Given that there are continuous families of representations this
greatly expands the state space.  This poses a very important
issue, which is that it may no longer
be possible to choose an inner product for the space
of diffeomorphism invariant states that renders it 
separable\footnote{There are delicate issues concerning
the treatment of the norm on states with high valence nodes, but these
may be resolved leading to separable HIlbert space.}.
This would be a disaster, which must be avoided if possible.

In fact, it is possible to avoid this disaster. To do this we work within the
Hilbert space whose basis is labeled by spin networks whose edges
are labeled only by pairs of ordinary spins $(j_{L},j_{R})$.  The
reason is that we will be implementing the Lorentzian signature
theory as long as we work in a space of states in which it is
possible to express, and solve, the operator forms of the reality
conditions, (\ref{Rab}) and (\ref{Sab}).  In this theory the 
kinematical theory differs from that of the Euclidean theory in that
every measure of three geometry, such as areas and volumes, has a
right value and a left value, which come from the corresponding 
labels on the states. This extension is the way that the spin network
theory can express the fact that it gives a kinematical description of
the complexification of geometry, in essence the complex part of
any function of the three metric is the difference between its left and right
value.  The reality conditions will, as we will see shortly, be 
expressed by conditions that require the left and right geometries to 
be equal.

Following the methods developed in \cite{lp1,sn1,gangof5} we are then 
free to impose the  condition that the states are
invariant under spatial
diffeomorphisms. Given the choice of kinematical inner
product on ${\cal H}^{gauge}$ defined by the $SU(2)_L \oplus SU(2)_R$
spin networks we construct in the usual way a unitary 
representation of the spatial diffeomorphism
group $Diff(\Sigma )$ on ${\cal H}^{gauge}$.
The gauge invariant states live in a subspace which is
called ${\cal H}^{diffeo}$.
These are by now standard constructions which were
done at the heuristic level in \cite{lp1,lp2,spain,sn1} and then
treated rigorously in \cite{reiner,chrisabhay,gangof5}.

Once ${\cal H}^{diffeo}$ is constructed there remain 
three more sets of
constraints to impose 
which are the Hamiltonian constraint 
${\cal C}(N)=0$ and the constraints $R^{ab}=0$ 
and $S^{ab}=0$ that determines
that the left and right handed fields define the same
metric geometry.

There are two ways we could handle the constraints
$R^{ab}=0$ and $S^{ab}=0$ that tie the left and right 
sectors to each other.  The 
orthodox 
Dirac method would require that, as these together make a second class 
set, that 
they be 
solved
explicitly and eliminated before the quantization.  One way to do this is to 
eliminate the 
right handed quantities 
$(A^{A'B'}_a , \pi^{aA'B'})$ together with the $SU(2)_R$
gauge freedom in favor of the left handed quantities.  This
would result in the Ashtekar formalism.  However, as the Hamiltonian is a 
constraint, there 
is a second option which can be tried, which may preserve the chiral
symmetry of the theory. This is to realize $R^{ab}$ as an 
operator equation
on physical states, so that we try to define and solve simultaneously
\f
\hat{R}^{ab} |\Psi \rangle =0 
\label{sym}
\ff
and 
\f
{\cal C} (N) |\Psi \rangle =0
\label{ham}
\ff
on states in ${\cal H}^{diffeo}$.

We then {\it define} the quantization of $S^{ab}$ by
\f
\hat{S}^{ab} \equiv [ \hat{\cal C}[N] , \hat{R}^{ab} ]
\ff
This is, of course, a formal expression that requires a
regularization procedure to specify completely.  It then
follows that physical states that satisfy (\ref{ham})
and (\ref{sym})  also satisfy
\f
\hat{S}^{ab} |\Psi \rangle =0
\ff

To see how this works, recall that standard constructions
give a normalizable basis for ${\cal H}^{gauge}$ in terms of
spin networks for the algebra $SU(2)_L \oplus SU(2)_R$.
The edges are labeled by pairs of spins $(j_L, j_R)$ and
the nodes are labeled by pairs of intertwiners
$(\mu_L , \mu_R )$

Using this basis it 
is easy to impose the condition (\ref{sym}) on states.
The reason is that $R^{ab}=0$ is equivalent to the requirement
that all area and volume observables constructed from $\pi^{aAB}$
and $\pi^{aA'B'}$ are equal.  For general states in the spin 
network basis, the areas and volumes constructed from $\pi^{aAB}$,
may be called the ``left quantum geometry.''  These will differ from those
constructed from $\pi^{aA'B'}$, which we may call the ``right handed
geometry''.  Classically $R^{ab}=0$ is equivalent to the statement
that the right handed areas and volumes are equal to the left handed
ones, for every region of the three manifold.  

The states in the spin network basis which are spanned by eigenstates
of left and right handed area and volume operators, such that the
eigenvalues of the left handed areas always equal the eigenvalues
of the righthanded areas, live in a subspace 
${\cal H}^{sym} \subset {\cal H}^{gauge}$ 
which is spanned by the subset of spin networks whose
labels satisfy
$j_L =j_R$ and $\mu_L = \mu_R $.  
Representations of $SU(2)_L \oplus SU(2)_R$ which
satisfy $j_L=j_R$  are called {\it balanced}.

Such representations have been employed by Barrett
and Crane in a proposal for a state sum model to
represent quantum general relativity\cite{johnlouis} and
have been studied recently in \cite{foam,kf}.  It is quite
interesting to find it arising also within the
Hamiltonian framework\footnote{An alternative approach to
deriving the Barrett-Crane balanced states as the quantization
of an action similar to (\ref{theaction} is given by
\cite{rl}.  In this approach one proceeds from the classical
action to the path integral directly by defining a natural
discretization of the $So(4)$ Plebanski action.}.

The restriction to balanced spin networks implements half
the reality conditions. The other half are, as we argued above,
automatically satisfied on states which are solutions to the
Hamiltonian constraints.  For the purposes of studying the
boundary theory, we need be concerned with one class of solutions
to the bulk Hamiltonian constraint, which are those that 
are derived from the Chern-Simons state\cite{kodama}. We thus,
now show that that state can be extended to the present case.

To do this we show that the loop transform of the 
$SU(2)_L \oplus SU(2)_R$ Chern-Simons state induces, as in
\cite{linking} a finite dimensional space of boundary states,
all of which satisfy the bulk Hamiltonian constraints.  These
are expressed in terms of arbitrary (but quantum deformed)
$SU(2)_L \oplus SU(2)_R$ spin networks.  The reality condition
(\ref{Rab}) is then implemented on this solution space by 
restricting the spin networks in the transform to the balanced
spin networks.  This restriction commutes with the imposition of
the constraints, so that the result also provides, by the formal
argument above, a solution to (\ref{Sab}).

The Chern-Simons state for $SU(2)_L \oplus SU(2)_R$ is given by
\f
\Psi_{CS}(A)= e^{{k' \over 4 \pi}
(S_{CS}[A_{AB}] - S_{CS}[A_{A'B'}])}
\ff
It is straightforward to show using the usual 
methods\cite{kodama,BGP} that
this solves independently both the left and right parts of the
Hamiltonian constraint (\ref{hams}).  In the spin network
basis, suitably quantum deformed\cite{qdef}, the corresponding
solutions space is given in the bulk by
\f
\Psi (\Gamma ) = \int DA e^{{k' \over 4 \pi}
(S_{CS}[A_{AB}] - S_{CS}[A_{A'B'}])} T[\Gamma ]
\label{solution}
\ff
here $k'={ 6 \pi \over G^2 \Lambda}$.  $\Gamma$ then refer to
$SU_{q}(2)_L \oplus SU_{q}(2)_R$ quantum spin networks, $q$ deformed with 
$q= e^{2\pi i/(k'+2)}$ and $ T[\Gamma ]$ is a suitably framed product
of traces of Wilson loops associated to $\Gamma$.  This defines
a finite dimensional state space, parameterized by boundary states
we will discuss shortly.  We may note that the fact that the cosmological 
constant is 
common to the left and right sector means that they have the same quantum 
deformation 
parameter.

The transform (\ref{solution}) defines a space of physical states, which
may be called ${\cal H}^{physical}$.  The restriction that defines
this may be stated as follows:a
functional of quantum deformed spin networks $\phi (\Gamma )$ is in 
${\cal H}^{physical}$ if it is invariant under the quantum recoupling
rules given in \cite{KL}.  The boundary theory then consists of
equivalence classes of quantum spin networks, under these recoupling
relations, which meet the boundary at a fixed set of punctures.

The reality condition (\ref{Sab}) can be imposed on the space of
states by requiring that the action of the left handed area operator,
defined from $\pi^{aAB}$ is equal to the action of the right handed
area operator $\pi^{aA'B'}$ for every two surface in the bulk.  These
actions are defined on the quantum deformed spin network states in
\cite{qdef}. The condition is solved for every surface when the
quantum deformed spin network states are restricted to balanced
spin networks.

To show that the Hamiltonian constraint has been solved in a way that
is consistent with the imposition of the reality conditions in this
form one must check that the restriction to balanced spin networks
commutes with the quantum recoupling relations, applied separately to
the left and right labels of the spin networks. This is 
straightforward, as one may use the recoupling relations to express
the equivalence classes in terms of trivalent spin networks, after 
which the imposition of the balanced conditions amount to the trivial
requirement that $j_{L}=j_{R}$ on all edges.  This means that, at 
least formally, the second reality condition (\ref{Sab}) is also
solved on this space of states.

We may now take into account the details of the construction of the
boundary theory.  The Chern-Simons state
state becomes a finite dimensional space of states, as
described in \cite{linking}, for each set of punctures on
the boundary.  These are given by the quantum deformed intertwiners on
the punctured boundary
The restriction to balanced representations
extends to the boundary, we also require that $k=k'$ so that
there is only one contribution to the cosmological constant.
One then constructs a space of physical states that has 
the form\cite{linking},
\f
{\cal H}^{phys}= \sum_{n} \sum_{j_1,...,j_n}  {\cal H}_{j_1,...,j_n}
\ff
where
\f
{\cal H}_{j_1,...,j_n} = {\cal V}_{j_1,...,j_n}^{balanced}
\subset
{\cal V}_{j_1,...,j_n}^L \otimes  {\cal V}_{j_1,...,j_n}^R .
\ff
Here ${\cal V}_{j_1,...,j_n}^{balanced}$ is the linear space of
balanced intertwiners in 
${\cal V}_{j_1,...,j_n}^L \otimes  {\cal V}_{j_1,...,j_n}^R$, which
is the space of conformal blocks for the
punctured sphere, for the $SU(2)_{L} \oplus SU(2)_{R}$
$WZW$ model.  By the balanced condition, the common
$j_i$'s label the punctures.  The sum extends up to
spins $j=k'$, because of the quantum deformation\cite{linking}.

The full set of physical observables for
the theory can be described in terms of operators on
${\cal H}^{phys}$ \cite{linking}.  Among them is the area
operator\cite{spain,vol1}, which is diagonal in the punctures and whose 
eigenvalues are given, in the limit of large $k'$ \cite{qdef}
by
\f
a[j_i]= \sum_i G\hbar \sqrt{j_i ( j_i +1)}
\ff
One then finds that the Bekenstein bound \cite{bek} is
satisfied, as
\f
\ln Dim \left ({\cal V}_{j_1,...,j_n}^L \otimes  {\cal V}_{j_1,...,j_n}^R
\right )  \leq c a[j_i]
\ff
with $c=2\sqrt{3}/ln(2)$.

Finally,to realize the dynamics we must implement the
hamiltonian (\ref{boundham}), which we recall is
a boundary term. The evolution
of the states according to a time defined on the boundary
by the function field $\tau  \in \partial {\cal M}$
is then given by the Schroedinger equation
\f
\imath \hbar {\delta \over \delta \tau} \Psi = 
\int_{\partial \Sigma} \tau \hat{h} \Psi
\ff
where the hamiltonian is given by an operator representing
(\ref{boundham}.)  The implementation of the hamiltonian as a quantum
operator on the space ${\cal H}^{phys}$ is a non-trivial problem,
which will represent another step in this program.  

Thus, as in the classical theory, the gauge invariance
splits up into a kinematical, linear part and a dynamical,
non-linear part, and the splitting affects both the bulk and
the boundary theory.  The linear kinematical part has to
do with the gauge invariance in the subgroup
$H=SU(2)_L \oplus SU(2)_R$, while the non-linear part has
to do with the coset $SP(4)/H$.  In the bulk the non-linear part of the gauge 
invariance 
turns out to be expressed precisely as the hamiltonian and diffeomorphism 
constraints of 
the theory.  In the boundary theory the linear part tells us that 
the theory has 
a complete 
holographic formulation given in terms of states and operators 
constructed from 
an 
ordinary conformal field theory
on the two dimensional spatial boundary.   The non-linear part of the gauge 
invariance, when extended to the boundary, gives rise to the hamiltonian, 
that generates physical time 
evolution.  This hamiltonian respects the preferred time slicing of the
boundary that was used for the construction of the boundary conditions
on the finite boundary.

Thus, we have found that general relativity with a cosmological constant has,
in the Lorentzian case, as well as the Euclidean case studied in 
\cite{linking}, an 
holographic formulation when expressed in terms of finite boundaries.  
The hope in subsequent work will be to extended this to the $N=8$ supersymmetric 
case and by doing so obtain results relevant for a 
holographic formulation of $\cal M$ theory.

\section*{ACKNOWLEDGEMENTS}

I am grateful to Abhay Ashtekar, Arivand Asok, Louis Crane,
Laurant Friedel,
Sameer Gupta, Kirill Krasnov,  
Fotini Markopoulou, Roger Penrose, Carlo Rovelli,
Edward Witten and especially Yi Ling for conversations
during this work.   This work was supported by
NSF grant PHY-9514240 to The Pennsylvania State
University and a gift from the
Jesse Phillips foundation.

\end{document}